%% file: main.tex
\documentclass[aps,prl,twocolumn,superscriptaddress,nofootinbib]{revtex4-2}
\usepackage{amsmath, bm}

\input{macros.tex}
\usepackage{graphicx}
\usepackage{xspace}
\usepackage{array}
\usepackage{float} %
\usepackage{subfigure}
\usepackage{comment} 
\usepackage{pdfpages}
 \usepackage{ifpdf}
 \ifpdf
 \usepackage[pdftex]{hyperref}
 \else
 \usepackage[hypertex]{hyperref}
 \fi

 \hypersetup{
   pdftitle={},%
   pdfauthor={},%
   pdfsubject={},%
   pdfkeywords={},%
   pdfstartview={},%
   bookmarksopen=true, breaklinks=true, debug=true, %
   colorlinks=true, linkcolor=red, citecolor=blue, urlcolor=blue
 }

\makeatletter						      
\AtBeginDocument{\let\LS@rot\@undefined}		      
\makeatother	
\begin{document}
\title{Impact of relativistic corrections to high-$P_T$ prompt-$\psip$ production at hadron colliders}

\author{Valerio Bertone}
\affiliation{Universit\'e Paris-Saclay, CEA, IRFU, 91191 Gif-sur-Yvette, France}

\author{Jean-Philippe Lansberg}
\affiliation{Universit\'e Paris-Saclay, CNRS, IJCLab, 91405 Orsay, France}

\author{Kate Lynch}
\affiliation{School of Physics, University College Dublin, Dublin 4, Ireland}
\affiliation{Universit\'e Paris-Saclay, CNRS, IJCLab, 91405 Orsay, France}

\date{\today}

\begin{abstract}
{We study relativistic corrections to prompt $\psip$ production at high \pT in hadron colliders. Our calculation employs leading-power factorization with Fragmentation Functions (FFs) computed in nonrelativistic QCD and evolved to next-to-leading-logarithmic accuracy. Relativistic corrections increase the cross section significantly for the gluon channel, but moderately for charm. We perform a full analysis of uncertainties. We observe a good agreement with both ATLAS and CMS cross sections without the need of higher-order color-octet contributions. The polar anisotropy is found to be close to CMS data, {partly due to charm fragmentation}.}
\end{abstract}

\maketitle

{\it Introduction --- }\input{introduction}

{\it NRQCD factorization and leading-power fragmentation --- }\input{LPXSEC}

\input{results}

{\it Conclusions --- }\input{Conclusion}

\begin{acknowledgments}
{\it Acknowledgments --- }
We are indebted to A.P.~Chen, Y.Q.~Ma, F.~Feng, Y. Jia, C.Y. Liu, J. Gao and D. Yang for their assistance in using their codes or results. We warmly thank M.~Fontannaz, M.~Nefedov, H.~Sazdjian,T.~Rabemananjara  for their numerous inputs and comments, as well as M.~Butensch\"on, M. Cacciari, L.~Carcedo Salgado, H.S.~Chung, S.~Fleming, C.~Flett, V. Kartvelishvili, R.~McNulty, C.F.~Qiao, H.S.~Shao, P.~Taels, C.~Van Hulse and T.~Zakareishvili for discussions.  
\end{acknowledgments}

{\it Funding --- }The research conducted in this publication was funded by the Irish Research Council under grant number GOIPG/2022/478. This project has also received funding from the Agence Nationale de la Recherche (ANR) via the grants ANR-20-CE31-0015 (``PrecisOnium''), ANR-24-CE31-7061-01 (``3DLeaP"), and via the IDEX Paris-Saclay ``Investissements d'Avenir" (ANR-11-IDEX-0003-01) through the GLUODYNAMICS project funded by the ``P2IO LabEx (ANR-10-LABX-0038)" and through the Joint PhD Programme of Universit\'e Paris-Saclay (ADI). This work  was also partly supported by the French CNRS via the IN2P3 projects ``GLUE@NLO" and ``QCDFactorisation@NLO".
\bibliographystyle{utphys}
\bibliography{bib}

\clearpage


\section*{Supplementary material}

# Supplementary material for "Impact of relativistic corrections to high-$P_T$ prompt-$\psi(2S)$ production at hadron colliders"

Valerio Bertone,[1] Jean-Philippe Lansberg,[2] and Kate Lynch[3,2]

[1]*Université Paris-Saclay, CEA, IRFU, 91191 Gif-sur-Yvette, France*
[2]*Université Paris-Saclay, CNRS, IJCLab, 91405 Orsay, France*
[3]*School of Physics, University College Dublin, Dublin 4, Ireland*
(Dated: October 7, 2025)

## I. FRAGMENTATION FUNCTION INITIAL CONDITIONS

Our expression for the unpolarized charm fragmentation function (FF) at $\mu_0$ is

$$D_{c/\psi(2S)}(z,\mu_{R_0},\mu_0) = \langle O_0^{\psi(2S)}(^3S_1^{[1]})\rangle \frac{\alpha_s^2(\mu_{R_0})}{m_c^3}\left(d^{(2,0)}_{c/Q\bar{Q}[n]}(z,\mu_{R_0},\mu_0) + \langle v^2\rangle d^{(2,2)}_{c/Q\bar{Q}[n]}(z,\mu_{R_0},\mu_0) + \alpha_s(\mu_{R_0})d^{(3,0)}_{c/Q\bar{Q}[n]}(z,\mu_{R_0},\mu_0)\right), \quad (1)$$

where

$$d^{(2,0)}_{c/Q\bar{Q}[n]}(z,\mu_{R_0},\mu_0) = \frac{16}{243}\frac{z(1-z)^2(16 - 32z + 72z^2 - 32z^3 + 5z^4)}{(2-z)^6}, \quad (2)$$

$$d^{(2,2)}_{c/Q\bar{Q}[n]}(z,\mu_{R_0},\mu_0) = \frac{8z(1-z)^2(-1344 + 5184z - 14416z^2 + 18176z^3 - 8924z^4 + 2092z^5 - 183z^6)}{2187(2-z)^8}, \quad (3)$$

and

$$d^{(3,0)}_{c/Q\bar{Q}[n]}(z,\mu_{R_0},\mu_0) = \left(\frac{1}{2\pi}\beta_0 \ln\frac{\mu_{R_0}^2}{4m_c^2}\right)\times d^{(2,0)}_{c/Q\bar{Q}[n]}(z) + f(z)\frac{2\pi}{9} + \frac{1}{2\pi}\ln\frac{\mu_0^2}{9m_c^2}\int_z^1\frac{dy}{y}P_{QQ}(y)d^{(2,0)}_{c\to\psi}(z/y,\mu_0). \quad (4)$$

Equation 2 is the $\alpha_s^2 v^0$ contribution to the charm FF and we use the result of Ref. [1]. Equation 3 is the $\alpha_s^2 v^2$ contribution to the charm FF and we use the result of Ref. [2]. Finally, Equation 4 is the $\alpha_s^3 v^0$ contribution to the charm FF and we use the expression from Ref. [3], where $f(z)$ is given in parametric form in Eq.(38).

Our expression for the unpolarized $\alpha_s^3 v^0$ gluon FF at $\mu_0$ is

$$D_{g/\psi(2S)}(z,\mu_0) = \langle O_0^{\psi(2S)}(^3S_1^{[1]})\rangle \frac{\alpha_s^3(\mu_{R_0})}{m_c^3}\left(d^{(3,0)}_{g/Q\bar{Q}[n]}(z,\mu_{R_0},\mu_0) + \langle v^2\rangle d^{(3,2)}_{g/Q\bar{Q}[n]}(z,\mu_{R_0},\mu_0)\right), \quad (5)$$

where

$$d^{(3,0)}_{g/Q\bar{Q}[n]}(z,\mu_{R_0},\mu_0) = \left[\frac{80\pi^3}{27(2N_c)}\left(C(z)I_{13}(z) + \sum_{i=0}^{11}C_i(z)L_i(z)\right)\right] + \left\{c_0(z)\ln\frac{\mu_0}{m_c} + c_1(z)\right\} \quad (6)$$

and

$$d^{(3,2)}_{g/Q\bar{Q}[n]}(z,\mu_{R_0},\mu_0) = \begin{cases} 0, & z=0, \\ b_2 z + \alpha_1 \ln(1-z)(1-z) + \beta_1 \ln^2(1-z)(1-z) \\ \quad + \ln(z)(\mu_1 z + \mu_2 z^2) + \nu_1 z \ln^2(z) + \sum_{k_2=1}^{4}\omega_{1k_2}z(1-z)^{k_2}, & 0 < z < 1, \\ b_2, & z=1. \end{cases} \quad (7)$$

Equation 6 is composed of two terms. The first term, in square brackets, is from $g \to \psi(2S)gg$ and we take the expression from Ref. [4], where $I_{13}$, $L$, and $C$ are given, respectively, in Eqns. (56), (58), and (A1). As noted in footnote 2 of the main manuscript, Ma in Ref. [4] employs a different convention for the LDME than we do: $\langle O_0^{\psi(2S)}(^3S_1^{[1]})\rangle = 2N_c \langle O_0^{\psi(2S)}(^3S_1^{[1]})\rangle_{\text{Ma}}$. To avoid confusion, we make explicit our absorption of a factor of $2N_c$ into the first term of $d^{(3,0)}_{g/Q\bar{Q}[n]}$. The second term of Eq. 6, in curly brackets, is from $g \to \psi(2S)c\bar{c}$ and was provided by Feng et al. (based on [5]). Results can be made available upon request to them.

Equation 7 gives the $\alpha_s^3 v^2$ corrections to the $g \to \psi(2S)gg$ contribution and we make use of the parametric form presented in the Appendix of [6], where $b_2$ is given in Eq.(B.1b) and constants $\alpha$, $\beta$, $\mu$, $\nu$, and $\omega$ are presented in Table 3 under column "$d_2(z)$".

## II. MELLIN MOMENTS AND $z$-DEPENDENCE OF FRAGMENTATION FUNCTIONS AT $\mu_0$ AND 100 GeV

Table I reports the $N^{\text{th}}$-Mellin moments of each of the $z$-dependent functions appearing in Eqns. 1 and 5. The moments for each are computed separately using $\mu_0 = \mu_{R_0}^g = \mu_{R_0}^c = 2m_c$ and $n_f = 3$. We compute the $N = 1, 2, 3, 4, 5, 6$ moments at $\mu_0$ and at $\mu = 100$ GeV. The $N = 1$ moment is the fragmentation probability, the ratio of the $N = 2$ to the $N = 1$ moments gives the average $z$, and at LHC energies, the cross section is sensitive to the $N \simeq 6$ moment (Ref. [6] uses 6.2). We separate the contributions to $d^{(3,0)}_{g/Q\bar{Q}[^3S_1^{[1]}]}$ into: (i) $g \to \psi(2S)gg$ and (ii) $g \to \psi(2S)c\bar{c}$ in the Table. The $g \to \psi(2S)c\bar{c}$ contribution is peaked at $z \to 0$, hence the large $N = 1$ moments. The moments of both the charm and gluon channels at $\mu = 100$ GeV highlight the non-negligible contribution coming from off-diagonal splittings ($P_{cg}$ and $P_{gc}$). This is particularly true for $P_{cg}$ (compare the size at 100 GeV of the (A) gluon-initiated to the (B) charm-initiated gluon moments).

TABLE I. Mellin moments of $z$-dependent functions appearing in Eqns. 1 and 5 with $\mu_0 = \mu_{R_0}^g = \mu_{R_0}^c = 2m_c$ and $n_f = 3$. Quantities (A) at 100 GeV arise from diagonal splitting kernels and (B) from off-diagonal splitting kernels. We separate $d^{(3,0)}_{g/Q\bar{Q}[^3S_1^{[1]}]}$ into (i) $g \to \psi(2S)gg$ and (ii) $g \to \psi(2S)c\bar{c}$. Each moment is multiplied by an overall factor of $10^4$.

|  | Gluon-initiated channel | | | Charm-initiated channel | | |
|---|---|---|---|---|---|---|
|  | $D_{c/\psi(2S)}(z, \mu_0) = 0$ | | | $D_{g/\psi(2S)}(z, \mu_0) = 0$ | | |
| (A) | $d^{(3,0)}_{g/Q\bar{Q}[^3S_1^{[1]}]}$ | | $d^{(3,2)}_{g/Q\bar{Q}[^3S_1^{[1]}]}$ | $d^{(2,0)}_{c/Q\bar{Q}[^3S_1^{[1]}]}$ | $d^{(3,0)}_{c/Q\bar{Q}[^3S_1^{[1]}]}$ | $d^{(2,2)}_{c/Q\bar{Q}[^3S_1^{[1]}]}$ |
|  | (i) | (ii) |  |  |  |  |
| $\mu = \mu_0$ | | | | | | |
| $N = 1$ | 8.29 | 148.92 | 20.33 | 81.61 | 60.12 | 20.85 |
| $N = 2$ | 3.67 | 27.29 | 18.53 | 50.36 | 50.53 | 23.36 |
| $N = 3$ | 2.13 | 8.84 | 16.16 | 34.38 | 38.68 | 20.67 |
| $N = 4$ | 1.42 | 3.83 | 14.22 | 24.90 | 29.86 | 17.43 |
| $N = 5$ | 1.02 | 1.98 | 12.68 | 18.77 | 23.47 | 14.55 |
| $N = 6$ | 0.78 | 1.15 | 11.45 | 14.58 | 18.80 | 12.16 |
| $\mu = 100$ GeV | | | | | | |
| $N = 1$ | 30.00 | 633.35 | 26.78 | 78.25 | 55.99 | 18.61 |
| $N = 2$ | 3.07 | 22.88 | 15.23 | 35.73 | 35.85 | 16.57 |
| $N = 3$ | 0.82 | 3.40 | 6.15 | 19.88 | 22.37 | 11.95 |
| $N = 4$ | 0.35 | 0.95 | 3.50 | 12.46 | 14.94 | 8.72 |
| $N = 5$ | 0.19 | 0.36 | 2.30 | 8.39 | 10.49 | 6.50 |
| $N = 6$ | 0.11 | 0.17 | 1.64 | 5.94 | 7.66 | 4.95 |
| (B) | $d^{(3,0)}_{c/Q\bar{Q}[^3S_1^{[1]}]}$ | | $d^{(3,2)}_{c/Q\bar{Q}[^3S_1^{[1]}]}$ | $d^{(2,0)}_{g/Q\bar{Q}[^3S_1^{[1]}]}$ | $d^{(3,0)}_{g/Q\bar{Q}[^3S_1^{[1]}]}$ | $d^{(2,2)}_{g/Q\bar{Q}[^3S_1^{[1]}]}$ |
| $\mu = 100$ GeV | | | | | | |
| $N = 1$ | 7.77 | 243.94 | -13.39 | 4.55 | 0.57 | -1.16 |
| $N = 2$ | 1.06 | 7.91 | 5.19 | 1.90 | 1.91 | 0.88 |
| $N = 3$ | 0.21 | 0.85 | 1.53 | 0.61 | 0.69 | 0.37 |
| $N = 4$ | 0.07 | 0.18 | 0.67 | 0.27 | 0.33 | 0.19 |
| $N = 5$ | 0.03 | 0.06 | 0.36 | 0.14 | 0.18 | 0.11 |
| $N = 6$ | 0.02 | 0.02 | 0.22 | 0.08 | 0.11 | 0.07 |

Figure 1 shows the fragmentation functions (Eqns. 1 and 5) with $\langle v^2_{\psi(2S)} \rangle$ = 0.00, 0.25, and 0.50 at initial scale $\mu_0$ and at $\mu = 100$ GeV for different scale choices. The scales are $(\mu_0, \mu^g_{R_0}, \mu^c_{R_0}) = (\xi_0 2m_c, \mu^g_{R_0} 2m_c, \mu^c_{R_0} m_c)$, where $\xi$ set to unity corresponds to the default scale choice. The large-$z$ region increases with increasing $\langle v^2_{\psi(2S)} \rangle$, this effect is moderate for the charm channel (green) but large for the gluon (gray-blue).

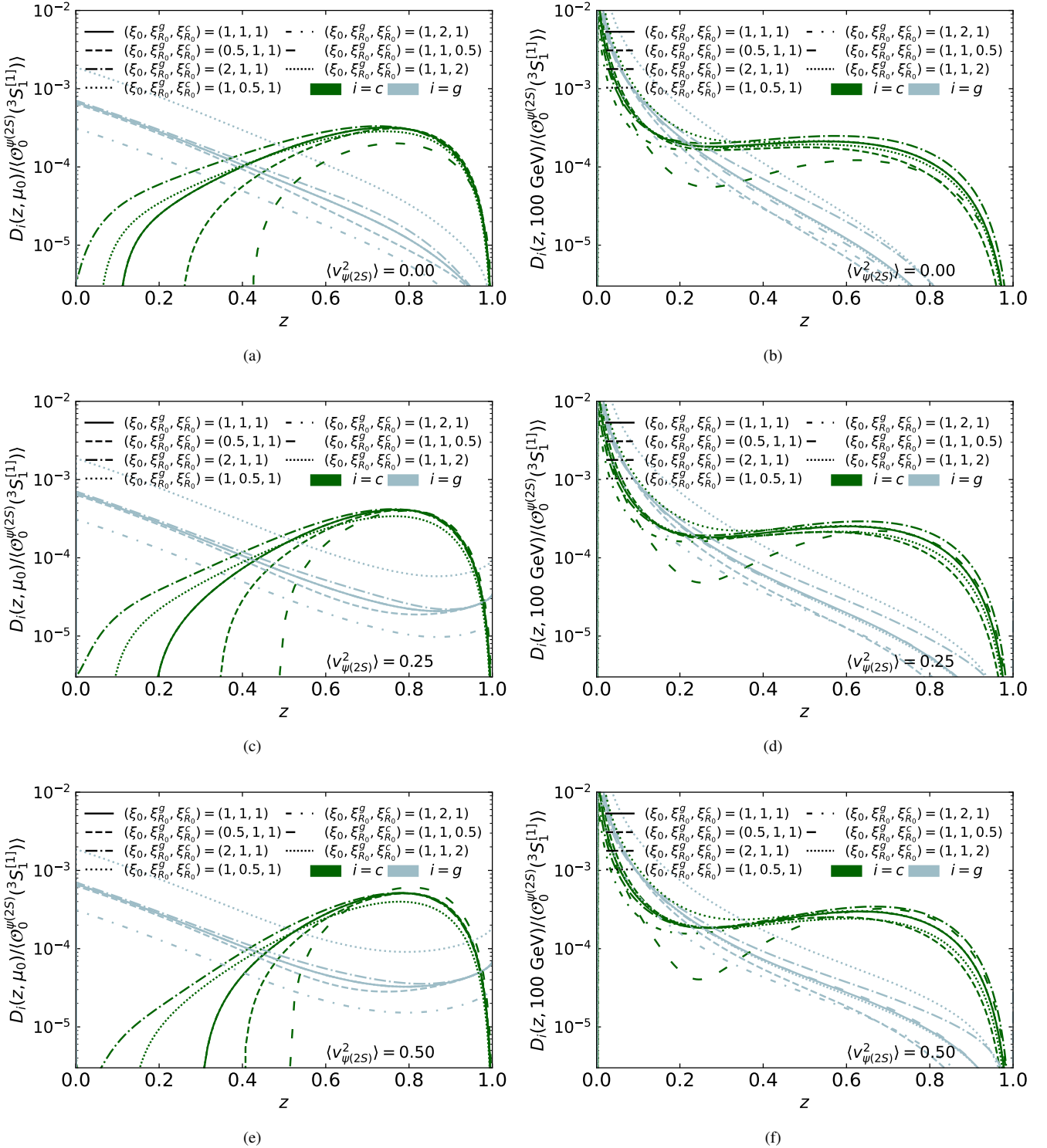

FIG. 1. Coupled charm (green) and gluon (gray-blue) fragmentation functions at (a,c,e) $\mu = \mu_0$ and (b,d,f) $\mu = 100$ GeV with (a,b) $\langle v^2_{\psi(2S)} \rangle$ = 0.00, (c,d) $\langle v^2_{\psi(2S)} \rangle$ = 0.25, and (e,f) $\langle v^2_{\psi(2S)} \rangle$ = 0.50 for each of the different scale choices indicated with different line styles.

## III. CROSS-SECTION-UNCERTAINTY BREAKDOWN AND DIFFERENTIAL CROSS-SECTION COMPARISON TO ATLAS DATA

Tables II and III report the $P_T$-differential cross section along with the breakdown of its total asymmetric uncertainty $\Delta\sigma_{\text{tot}}^{\pm}$ coming from each of the scale variations and the PDF uncertainty for, respectively, the default ($|R(0)_{\psi(2S)}|^2 = 0.53$ GeV$^3$, $\langle v^2_{\psi(2S)} \rangle = 0.5$) and alternate ($|R(0)_{\psi(2S)}|^2 = 0.93$ GeV$^3$, $\langle v^2_{\psi(2S)} \rangle = 0.25$) scenario. The tabulated results correspond to the binned cross sections in Fig. 1(c) of the main manuscript. At next-to-leading order in $\alpha_s$, the charm FF renormalization scale dependence $\mu_{R_0}^c$ is non-trivial and, in the alternate scenario, the central scale choice gives the largest cross section, hence $\Delta\sigma_{\mu_{R_0}^c}^+ = 0$.

TABLE II. $P_T$-differential cross section results and uncertainty breakdown for default scenario ($|R(0)_{\psi(2S)}|^2 = 0.53$ GeV$^3$, $\langle v^2_{\psi(2S)} \rangle = 0.5$) [See to the upper panel of Fig. 1(c) of main manuscript].

| $P_T$ | $d\sigma/dP_T \times \mathcal{B}$ | $\Delta\sigma_{\text{tot}}$ | $\Delta\sigma_{\mu_{R_0}^g}$ | $\Delta\sigma_{\mu_0}$ | $\Delta\sigma_{\mu_R}$ | $\Delta\sigma_{\mu_F}$ | $\Delta\sigma_{\mu_f}$ | $\Delta\sigma_{\mu_{R_0}^c}$ | $\Delta\sigma_{\text{PDF}}$ |
|---|---|---|---|---|---|---|---|---|---|
| GeV | nb/GeV | nb/GeV | nb/GeV | nb/GeV | nb/GeV | nb/GeV | nb/GeV | nb/GeV | nb/GeV |
| 20–22 | $4.6 \times 10^3$ | $+6.2\times10^3$ $-2.6\times10^3$ | $+5.8\times10^3$ $-1.7\times10^3$ | $+2.1\times10^3$ $-1.7\times10^3$ | $+8.7\times10^2$ $-7.3\times10^2$ | $+5.5\times10^2$ $-4.1\times10^2$ | $+2.9\times10^2$ $-4.0\times10^2$ | $+1.5\times10^2$ $-2.8\times10^2$ | $+1.2\times10^2$ $-1.3\times10^2$ |
| 22–24 | $3.0 \times 10^3$ | $+4.0\times10^3$ $-1.7\times10^3$ | $+3.7\times10^3$ $-1.1\times10^3$ | $+1.4\times10^3$ $-1.1\times10^3$ | $+5.4\times10^2$ $-4.6\times10^2$ | $+3.5\times10^2$ $-2.6\times10^2$ | $+1.8\times10^2$ $-2.3\times10^2$ | $+1.0\times10^2$ $-1.9\times10^2$ | $+7.4\times10^1$ $-7.8\times10^1$ |
| 24–26 | $2.0 \times 10^3$ | $+2.7\times10^3$ $-1.1\times10^3$ | $+2.4\times10^3$ $-7.3\times10^2$ | $+9.3\times10^2$ $-7.3\times10^2$ | $+3.5\times10^2$ $-3.0\times10^2$ | $+2.3\times10^2$ $-1.7\times10^2$ | $+1.1\times10^2$ $-1.4\times10^2$ | $+7.1\times10^1$ $-1.3\times10^2$ | $+4.8\times10^1$ $-5.1\times10^1$ |
| 26–28 | $1.4 \times 10^3$ | $+1.8\times10^3$ $-7.5\times10^2$ | $+1.7\times10^3$ $-5.0\times10^2$ | $+6.4\times10^2$ $-5.0\times10^2$ | $+2.4\times10^2$ $-2.0\times10^2$ | $+1.6\times10^2$ $-1.1\times10^2$ | $+7.2\times10^1$ $-8.5\times10^1$ | $+5.0\times10^1$ $-9.2\times10^1$ | $+3.2\times10^1$ $-3.4\times10^1$ |
| 28–30 | $9.6 \times 10^2$ | $+1.3\times10^3$ $-5.3\times10^2$ | $+1.2\times10^3$ $-3.5\times10^2$ | $+4.6\times10^2$ $-3.5\times10^2$ | $+1.6\times10^2$ $-1.4\times10^2$ | $+1.1\times10^2$ $-8.0\times10^1$ | $+4.8\times10^1$ $-5.5\times10^1$ | $+3.7\times10^1$ $-6.6\times10^1$ | $+2.2\times10^1$ $-2.4\times10^1$ |
| 30–35 | $6.2 \times 10^2$ | $+8.0\times10^2$ $-3.4\times10^2$ | $+7.4\times10^2$ $-2.2\times10^2$ | $+2.9\times10^2$ $-2.3\times10^2$ | $+9.9\times10^1$ $-8.5\times10^1$ | $+6.7\times10^1$ $-5.0\times10^1$ | $+2.8\times10^1$ $-3.2\times10^1$ | $+2.5\times10^1$ $-4.3\times10^1$ | $+1.4\times10^1$ $-1.5\times10^1$ |
| 35–40 | $2.9 \times 10^2$ | $+3.7\times10^2$ $-1.6\times10^2$ | $+3.4\times10^2$ $-1.0\times10^2$ | $+1.4\times10^2$ $-1.1\times10^2$ | $+4.4\times10^1$ $-3.9\times10^1$ | $+3.0\times10^1$ $-2.3\times10^1$ | $+1.0\times10^1$ $-1.3\times10^1$ | $+1.3\times10^1$ $-2.2\times10^1$ | $+6.1\times10^0$ $-6.7\times10^0$ |
| 40–45 | $1.6 \times 10^2$ | $+2.0\times10^2$ $-8.4\times10^1$ | $+1.8\times10^2$ $-5.4\times10^1$ | $+7.5\times10^1$ $-5.8\times10^1$ | $+2.2\times10^1$ $-2.0\times10^1$ | $+1.6\times10^1$ $-1.2\times10^1$ | $+4.6\times10^0$ $-5.9\times10^0$ | $+7.2\times10^0$ $-1.2\times10^1$ | $+3.1\times10^0$ $-3.4\times10^0$ |
| 45–50 | $8.7 \times 10^1$ | $+1.1\times10^2$ $-4.6\times10^1$ | $+9.9\times10^1$ $-2.9\times10^1$ | $+4.2\times10^1$ $-3.2\times10^1$ | $+1.2\times10^1$ $-1.1\times10^1$ | $+8.4\times10^0$ $-6.5\times10^0$ | $+2.1\times10^0$ $-2.6\times10^0$ | $+4.2\times10^0$ $-6.9\times10^0$ | $+1.6\times10^0$ $-1.8\times10^0$ |
| 50–60 | $4.4 \times 10^1$ | $+5.4\times10^1$ $-2.3\times10^1$ | $+4.9\times10^1$ $-1.5\times10^1$ | $+2.1\times10^1$ $-1.6\times10^1$ | $+5.9\times10^0$ $-5.4\times10^0$ | $+4.1\times10^0$ $-3.2\times10^0$ | $+8.7\times10^{-1}$ $-9.9\times10^{-1}$ | $+2.2\times10^0$ $-3.6\times10^0$ | $+7.9\times10^{-1}$ $-8.9\times10^{-1}$ |
| 60–70 | $1.8 \times 10^1$ | $+2.2\times10^1$ $-9.4\times10^0$ | $+2.0\times10^1$ $-5.8\times10^0$ | $+8.7\times10^0$ $-6.7\times10^0$ | $+2.2\times10^0$ $-2.1\times10^0$ | $+1.6\times10^0$ $-1.3\times10^0$ | $+2.7\times10^{-1}$ $-2.8\times10^{-1}$ | $+9.6\times10^{-1}$ $-1.5\times10^0$ | $+3.1\times10^{-1}$ $-3.5\times10^{-1}$ |
| 70–80 | $8.3 \times 10^0$ | $+9.8\times10^0$ $-4.3\times10^0$ | $+8.8\times10^0$ $-2.6\times10^0$ | $+4.0\times10^0$ $-3.1\times10^0$ | $+9.9\times10^{-1}$ $-9.3\times10^{-1}$ | $+7.0\times10^{-1}$ $-5.6\times10^{-1}$ | $+9.1\times10^{-2}$ $-8.4\times10^{-2}$ | $+4.6\times10^{-1}$ $-7.3\times10^{-1}$ | $+1.4\times10^{-1}$ $-1.5\times10^{-1}$ |
| 80–90 | $4.2 \times 10^0$ | $+4.8\times10^0$ $-2.1\times10^0$ | $+4.4\times10^0$ $-1.3\times10^0$ | $+2.0\times10^0$ $-1.5\times10^0$ | $+4.8\times10^{-1}$ $-4.5\times10^{-1}$ | $+3.4\times10^{-1}$ $-2.7\times10^{-1}$ | $+2.5\times10^{-2}$ $-2.4\times10^{-2}$ | $+2.4\times10^{-1}$ $-3.8\times10^{-1}$ | $+6.8\times10^{-2}$ $-7.5\times10^{-2}$ |
| 90–100 | $2.2 \times 10^0$ | $+2.6\times10^0$ $-1.1\times10^0$ | $+2.3\times10^0$ $-6.9\times10^{-1}$ | $+1.1\times10^0$ $-8.4\times10^{-1}$ | $+2.5\times10^{-1}$ $-2.4\times10^{-1}$ | $+1.8\times10^{-1}$ $-1.5\times10^{-1}$ | $+3.8\times10^{-3}$ $-5.9\times10^{-3}$ | $+1.4\times10^{-1}$ $-2.1\times10^{-1}$ | $+3.7\times10^{-2}$ $-4.0\times10^{-2}$ |
| 100–120 | $1.0 \times 10^0$ | $+1.2\times10^0$ $-5.2\times10^{-1}$ | $+1.0\times10^0$ $-3.1\times10^{-1}$ | $+5.0\times10^{-1}$ $-3.8\times10^{-1}$ | $+1.1\times10^{-1}$ $-1.1\times10^{-1}$ | $+7.9\times10^{-2}$ $-6.6\times10^{-2}$ | $+1.5\times10^{-3}$ $-1.9\times10^{-3}$ | $+6.6\times10^{-2}$ $-9.7\times10^{-2}$ | $+1.7\times10^{-2}$ $-1.8\times10^{-2}$ |
| 120–140 | $3.9 \times 10^{-1}$ | $+4.3\times10^{-1}$ $-2.0\times10^{-1}$ | $+3.9\times10^{-1}$ $-1.2\times10^{-1}$ | $+1.9\times10^{-1}$ $-1.5\times10^{-1}$ | $+4.1\times10^{-2}$ $-3.9\times10^{-2}$ | $+2.9\times10^{-2}$ $-2.4\times10^{-2}$ | $+3.2\times10^{-3}$ $-1.8\times10^{-3}$ | $+2.7\times10^{-2}$ $-3.8\times10^{-2}$ | $+6.9\times10^{-3}$ $-7.0\times10^{-3}$ |

TABLE III. $P_T$-differential cross section results and uncertainty breakdown for alternate scenario ($|R(0)_{\psi(2S)}|^2 = 0.93$ GeV$^3$, $\langle v^2_{\psi(2S)}\rangle = 0.25$) [See the lower panel of Fig. 1(c) of main manuscript].

| $P_T$ | $d\sigma/dP_T \times \mathcal{B}$ | $\Delta\sigma_{\text{tot}}$ | $\Delta\sigma_{\mu^g_{R_0}}$ | $\Delta\sigma_{\mu_0}$ | $\Delta\sigma_{\mu_R}$ | $\Delta\sigma_{\mu_F}$ | $\Delta\sigma_{\mu_f}$ | $\Delta\sigma_{\mu^c_{R_0}}$ | $\Delta\sigma_{\text{PDF}}$ |
|---|---|---|---|---|---|---|---|---|---|
| GeV | nb/GeV | nb/GeV | nb/GeV | nb/GeV | nb/GeV | nb/GeV | nb/GeV | nb/GeV | nb/GeV |
| 20–22 | $5.7 \times 10^3$ | $+7.6\times10^3$ / $-3.2\times10^3$ | $+6.9\times10^3$ / $-2.1\times10^3$ | $+2.8\times10^3$ / $-2.1\times10^3$ | $+1.1\times10^3$ / $-9.0\times10^2$ | $+6.4\times10^2$ / $-4.8\times10^2$ | $+3.3\times10^2$ / $-4.7\times10^2$ | $+0$ / $-3.2\times10^2$ | $+1.5\times10^2$ / $-1.5\times10^2$ |
| 22–24 | $3.7 \times 10^3$ | $+4.8\times10^3$ / $-2.0\times10^3$ | $+4.4\times10^3$ / $-1.3\times10^3$ | $+1.8\times10^3$ / $-1.4\times10^3$ | $+6.6\times10^2$ / $-5.6\times10^2$ | $+4.1\times10^2$ / $-3.0\times10^2$ | $+2.0\times10^2$ / $-2.7\times10^2$ | $+0$ / $-2.1\times10^2$ | $+9.1\times10^1$ / $-9.7\times10^1$ |
| 24–26 | $2.5 \times 10^3$ | $+3.2\times10^3$ / $-1.3\times10^3$ | $+2.9\times10^3$ / $-8.6\times10^2$ | $+1.2\times10^3$ / $-9.2\times10^2$ | $+4.2\times10^2$ / $-3.6\times10^2$ | $+2.7\times10^2$ / $-2.0\times10^2$ | $+1.3\times10^2$ / $-1.6\times10^2$ | $+0$ / $-1.5\times10^2$ | $+5.9\times10^1$ / $-6.3\times10^1$ |
| 26–28 | $1.7 \times 10^3$ | $+2.2\times10^3$ / $-9.2\times10^2$ | $+2.0\times10^3$ / $-5.9\times10^2$ | $+8.4\times10^2$ / $-6.3\times10^2$ | $+2.8\times10^2$ / $-2.4\times10^2$ | $+1.8\times10^2$ / $-1.3\times10^2$ | $+8.2\times10^1$ / $-9.9\times10^1$ | $+0$ / $-1.0\times10^2$ | $+3.9\times10^1$ / $-4.2\times10^1$ |
| 28–30 | $1.2 \times 10^3$ | $+1.5\times10^3$ / $-6.4\times10^2$ | $+1.4\times10^3$ / $-4.1\times10^2$ | $+5.9\times10^2$ / $-4.5\times10^2$ | $+1.9\times10^2$ / $-1.7\times10^2$ | $+1.3\times10^2$ / $-9.1\times10^1$ | $+5.5\times10^1$ / $-6.3\times10^1$ | $+0$ / $-7.5\times10^1$ | $+2.7\times10^1$ / $-2.9\times10^1$ |
| 30–35 | $7.8 \times 10^2$ | $+9.6\times10^2$ / $-4.1\times10^2$ | $+8.7\times10^2$ / $-2.6\times10^2$ | $+3.8\times10^2$ / $-2.9\times10^2$ | $+1.2\times10^2$ / $-1.0\times10^2$ | $+7.9\times10^1$ / $-5.6\times10^1$ | $+3.2\times10^1$ / $-3.7\times10^1$ | $+0$ / $-4.9\times10^1$ | $+1.7\times10^1$ / $-1.8\times10^1$ |
| 35–40 | $3.7 \times 10^2$ | $+4.4\times10^2$ / $-1.9\times10^2$ | $+4.0\times10^2$ / $-1.2\times10^2$ | $+1.8\times10^2$ / $-1.4\times10^2$ | $+5.2\times10^1$ / $-4.8\times10^1$ | $+3.5\times10^1$ / $-2.6\times10^1$ | $+1.2\times10^1$ / $-1.5\times10^1$ | $+0$ / $-2.5\times10^1$ | $+7.6\times10^0$ / $-8.4\times10^0$ |
| 40–45 | $2.0 \times 10^2$ | $+2.3\times10^2$ / $-1.0\times10^2$ | $+2.1\times10^2$ / $-6.3\times10^1$ | $+9.6\times10^1$ / $-7.3\times10^1$ | $+2.6\times10^1$ / $-2.5\times10^1$ | $+1.8\times10^1$ / $-1.4\times10^1$ | $+5.5\times10^0$ / $-6.8\times10^0$ | $+0$ / $-1.4\times10^1$ | $+3.9\times10^0$ / $-4.3\times10^0$ |
| 45–50 | $1.1 \times 10^2$ | $+1.3\times10^2$ / $-5.6\times10^1$ | $+1.1\times10^2$ / $-3.4\times10^1$ | $+5.4\times10^1$ / $-4.0\times10^1$ | $+1.4\times10^1$ / $-1.3\times10^1$ | $+9.8\times10^0$ / $-7.4\times10^0$ | $+2.6\times10^0$ / $-3.0\times10^0$ | $+0$ / $-7.9\times10^0$ | $+2.0\times10^0$ / $-2.3\times10^0$ |
| 50–60 | $5.6 \times 10^1$ | $+6.4\times10^1$ / $-2.8\times10^1$ | $+5.7\times10^1$ / $-1.7\times10^1$ | $+2.7\times10^1$ / $-2.1\times10^1$ | $+6.9\times10^0$ / $-6.5\times10^0$ | $+4.8\times10^0$ / $-3.7\times10^0$ | $+1.1\times10^0$ / $-1.2\times10^0$ | $+0$ / $-4.1\times10^0$ | $+9.9\times10^{-1}$ / $-1.1\times10^0$ |
| 60–70 | $2.3 \times 10^1$ | $+2.5\times10^1$ / $-1.1\times10^1$ | $+2.2\times10^1$ / $-6.7\times10^0$ | $+1.1\times10^1$ / $-8.3\times10^0$ | $+2.6\times10^0$ / $-2.5\times10^0$ | $+1.8\times10^0$ / $-1.4\times10^0$ | $+3.6\times10^{-1}$ / $-3.3\times10^{-1}$ | $+0$ / $-1.7\times10^0$ | $+3.9\times10^{-1}$ / $-4.3\times10^{-1}$ |
| 70–80 | $1.0 \times 10^1$ | $+1.1\times10^1$ / $-5.1\times10^0$ | $+1.0\times10^1$ / $-3.0\times10^0$ | $+5.1\times10^0$ / $-3.8\times10^0$ | $+1.2\times10^0$ / $-1.1\times10^0$ | $+8.2\times10^{-1}$ / $-6.3\times10^{-1}$ | $+1.3\times10^{-1}$ / $-9.6\times10^{-2}$ | $+0$ / $-8.3\times10^{-1}$ | $+1.7\times10^{-1}$ / $-1.9\times10^{-1}$ |
| 80–90 | $5.2 \times 10^0$ | $+5.6\times10^0$ / $-2.5\times10^0$ | $+5.0\times10^0$ / $-1.5\times10^0$ | $+2.6\times10^0$ / $-1.9\times10^0$ | $+5.6\times10^{-1}$ / $-5.5\times10^{-1}$ | $+3.9\times10^{-1}$ / $-3.1\times10^{-1}$ | $+3.9\times10^{-2}$ / $-2.9\times10^{-2}$ | $+0$ / $-4.3\times10^{-1}$ | $+8.7\times10^{-2}$ / $-9.4\times10^{-2}$ |
| 90–100 | $2.8 \times 10^0$ | $+3.0\times10^0$ / $-1.4\times10^0$ | $+2.6\times10^0$ / $-7.8\times10^{-1}$ | $+1.4\times10^0$ / $-1.0\times10^0$ | $+2.9\times10^{-1}$ / $-2.9\times10^{-1}$ | $+2.1\times10^{-1}$ / $-1.6\times10^{-1}$ | $+8.8\times10^{-3}$ / $-8.0\times10^{-3}$ | $+0$ / $-2.4\times10^{-1}$ | $+4.7\times10^{-2}$ / $-5.0\times10^{-2}$ |
| 100–120 | $1.3 \times 10^0$ | $+1.3\times10^0$ / $-6.2\times10^{-1}$ | $+1.2\times10^0$ / $-3.5\times10^{-1}$ | $+6.3\times10^{-1}$ / $-4.7\times10^{-1}$ | $+1.3\times10^{-1}$ / $-1.3\times10^{-1}$ | $+9.2\times10^{-2}$ / $-7.5\times10^{-2}$ | $+1.1\times10^{-3}$ / $-4.1\times10^{-4}$ | $+0$ / $-1.1\times10^{-1}$ | $+2.2\times10^{-2}$ / $-2.3\times10^{-2}$ |
| 120–140 | $4.9 \times 10^{-1}$ | $+5.0\times10^{-1}$ / $-2.3\times10^{-1}$ | $+4.3\times10^{-1}$ / $-1.3\times10^{-1}$ | $+2.4\times10^{-1}$ / $-1.8\times10^{-1}$ | $+4.8\times10^{-2}$ / $-4.7\times10^{-2}$ | $+3.4\times10^{-2}$ / $-2.8\times10^{-2}$ | $+3.3\times10^{-3}$ / $-1.5\times10^{-3}$ | $+0$ / $-4.4\times10^{-2}$ | $+8.8\times10^{-3}$ / $-8.9\times10^{-3}$ |

Figure 2 compares prompt ATLAS $\psi(2S)$ $P_T$-differential cross section data [7] in three bins of rapidity at $\sqrt{s}$ = 13 TeV to our results for coupled $c$ & $g$ channels using $|R(0)_{\psi(2S)}|^2$ = 0.53 GeV$^3$, $\langle v^2_{\psi(2S)}\rangle$ = 0.50 (a) and $|R(0)_{\psi(2S)}|^2$ = 0.93 GeV$^3$, $\langle v^2_{\psi(2S)}\rangle$ = 0.25 (b).

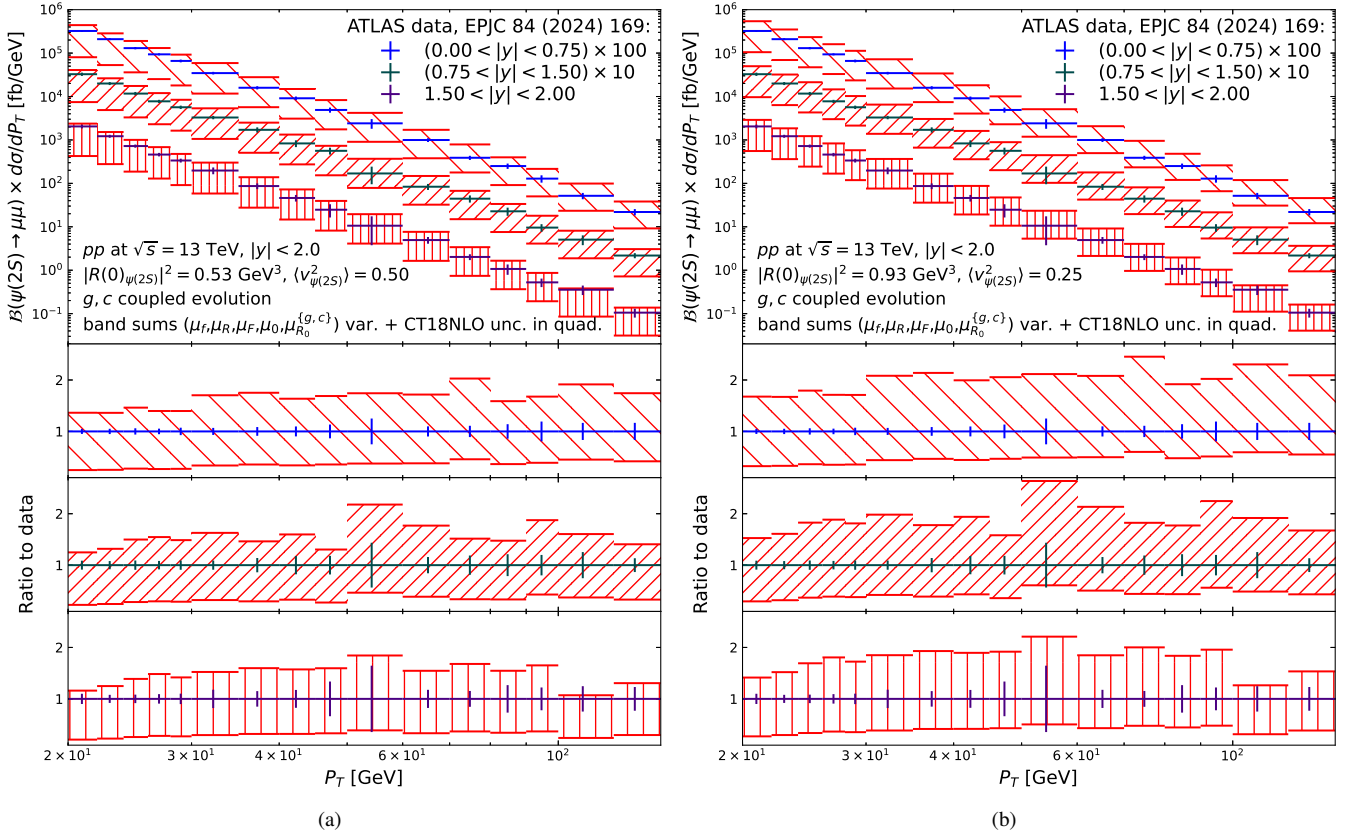

(a)

(b)

FIG. 2. Prompt $\psi(2S)$ ATLAS data [7] in three bins of rapidity: $0.00 < |y| < 0.75$ (red), $0.75 < |y| < 1.50$ (green), and $1.50 < |y| < 2.00$ (blue), compared to coupled charm and gluon evolution results (hatched bands) using (a) $|R_{\psi(2S)}(0)|^2$ = 0.53 GeV$^3$, $\langle v^2_{\psi(2S)}\rangle$ = 0.50 and (b) $|R_{\psi(2S)}(0)|^2$ = 0.93 GeV$^3$, $\langle v^2_{\psi(2S)}\rangle$ = 0.25.

## IV. CDF DATA COMPARISON

Figure 3 compares prompt CDF $\psi(2S)$ $P_T$-differential cross section at $\sqrt{s}$ = 1.96 TeV to our results for coupled $c$ & $g$ (hatched red), $c$ (solid green), and $g$ (solid light blue) channels using $|R(0)_{\psi(2S)}|^2 = 0.53$ GeV$^3$, $\langle v^2_{\psi(2S)} \rangle = 0.5$.

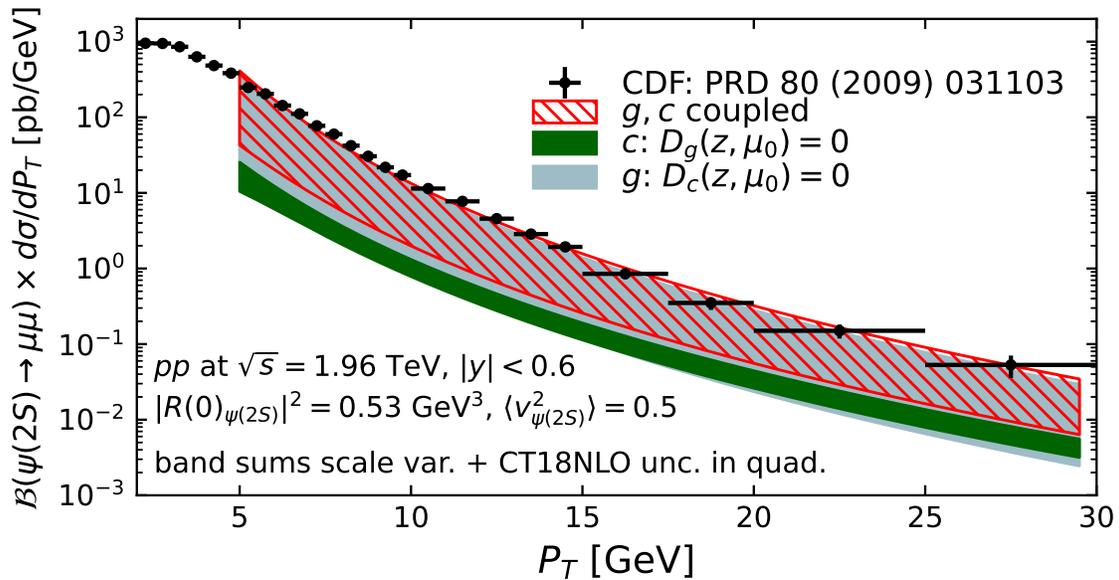

FIG. 3. Run-2 prompt CDF $\psi(2S)$ $P_T$-differential cross section data [8] at $\sqrt{s}$ = 1.96 TeV compared to our results for coupled $c$ & $g$ (hatched red), $c$ (solid green), and $g$ (solid light blue) channels.

 
\end{document}

%% file: macros.tex
\newcommand{\ce}[1]{Eq.~(\ref{#1})}
\newcommand{\cf}[1]{{Fig.~\ref{#1}}}

\DeclareMathAlphabet{\pazocal}{OMS}{zplm}{m}{n}
\newcommand{\Q}{\ensuremath{\pazocal{Q}}}
\newcommand{\QQ}{\ensuremath{Q\bar{Q}}}

\usepackage{color}
\usepackage[normalem]{ulem}
\usepackage{cancel}
\usepackage{txfonts}

\newcommand{\pT}{\ensuremath{P_T}\xspace}
\newcommand{\jpsi}{\ensuremath{J/\psi}\xspace}
\newcommand{\psip}{\ensuremath{\psi(2S\!)}\xspace}

%% file: introduction.tex
Quarkonia ($\Q$), made up of two heavy quarks, are probably the simplest QCD bound states in nature. 
However, their production, which involves strong interactions at both short and long distances when the heavy-quark pair is produced and hadronizes, respectively, remains poorly understood.
Currently, none of the proposed production mechanisms~\cite{Chang:1979nn,Berger:1980ni,Baier:1981uk,Bodwin:1994jh,Ma:2016exq,Ma:2017xno} describes the wide variety of existing measurements~\cite{Lansberg:2019adr}. To advance our understanding, it is important to find a good trade-off between the complexity of theoretical computations and of experimental measurements. In this context, high-$P_T$ prompt production of $\psip$ is optimal, as it is not polluted by decays of other quarkonia and is simple to measure. Its measurement at the Tevatron by CDF in the 90's~\cite{CDF:1997ykw}, then confirmed in the 2000's~\cite{CDF:2009kwm}, uncovered an anomaly called the \psip ``surplus''. This led to the introduction~\cite{Braaten:1994vv} of nonrelativistic QCD (NRQCD)~\cite{Bodwin:1994jh}, which results in a rigorous factorization between Short-Distance Coefficients (SDCs) and Long-Distance Matrix Elements (LDMEs).

At the LHC, the $P_T$ reach of $\psip$ measurements is four times~\cite{CMS:2017dju,ATLAS:2023qnh} larger than at Tevatron and spans a domain where Leading-Power (LP) factorization, written in terms of single-parton Fragmentation Functions (FFs), dominates~\cite{Braaten:1993rw,Braaten:1994xb,Cacciari:1994dr,Bodwin:2014gia,Ma:2014svb}. In fact, LP contributions to $d\sigma/dP_T^{2}$ scale as $P_T^{-4}$, while others scale at most as $P_T^{-6}$. At LP, $\Q$ production in NRQCD is further simplified (and made more precise) with a further factorization of the SDCs into perturbatively calculable FFs and the usual partonic cross sections deriving from collinear factorization~\cite{CTEQ:1993hwr}. Since FFs are naturally defined at scales of the order of the $\Q$ mass, $M$, if $P_T\gg M$ large logarithms of $P_T/M$ can be resummed through collinear evolution. Using LP factorization at high $P_T$~\cite{Kang:2014pya,Ma:2014svb} also prevents the appearance of large radiative QCD corrections in Fixed-Order (FO) computations of partonic cross sections~\cite{Kramer:1995nb,Gong:2008sn,Lansberg:2009db,Gong:2012ah,Lansberg:2013qka,Lansberg:2014swa,Flore:2020jau,Feng:2025btt}.

Despite a reduced $P_T$ reach, the case of $\psip$ is significantly
better than that of $J/\psi$~\cite{ATLAS:2023qnh}. Indeed, since the pioneering CDF {studies}~\cite{CDF:1997uzj}, it is known that about a third of the $J/\psi$ yield comes from $\chi_c$ feed-down. As $\chi_c$ measurements~\cite{ATLAS:2014ala} barely reach $P_T=25$~GeV, this amounts to a 30\% systematic uncertainty because, at larger $P_T$, NRQCD features instabilities that require specific treatments~\cite{Chung:2023ext,Chen:2023gsu,Chung:2024jfk} beyond LP.

In this Letter, we present the first impact study of relativistic corrections to prompt hadroproduction of $\psip$ at large $P_T$. We include LP fragmentation from charm quarks $c$ and gluons $g$, along with their coupled evolution up to {next-to-leading logarithmic} (NLL) accuracy (resumming terms of order \(\alpha_s^{\,n+1} \ln^n(P_T/M)\)), SDCs up to next-to-leading order (NLO), i.e., $\mathcal{O}(\alpha_s^3)$, and relativistic corrections up to \(\mathcal{O}(v^2)\), where \(v\) is the velocity of the $c$ in the rest frame of the pair. Evolution is performed in the so-called Variable-Flavor-Number Scheme (VFNS), which accounts for heavy-flavor--threshold crossing. We  account for NLO corrections ($\mathcal{O}(\alpha^3_s)$) to the charm FF and for the $g \to \psip c\bar c$ contribution of the gluon FF. We compare our results (cross sections and polar anisotropy) to the latest prompt ATLAS and CMS data.

%% file: LPXSEC.tex
The $P_T$-differential cross section for the inclusive production of a $\pazocal{Q}$ in a collision between hadrons $h_A$ and $h_B$ can be written, up to {corrections {scaling} like} $M^2/P_T^2$~\cite{Collins:1981uk}, as
\begin{align}\label{eq:LP}
   & \frac{d\sigma_{h_Ah_B \to \Q X}}{dP_{T}} = \sum_{i,j,k} \int dx_i dx_j dz \, f_{i/h_A}(x_i;\mu_f)   f_{j/h_B}(x_j;\mu_f)  \nonumber  \\
   &\times \frac{d\hat{\sigma}_{i j \to k X }}{dP_{T}} (x_i,x_j,z,P_T;\mu_f,\mu_R,\mu_F) D_{k/\Q}(z;\mu_F), 
\end{align}
where $x_i$, $x_j$ are the momentum fractions of the {incoming} partons {$i$ and $j$} relative to the {parent} hadrons, $z$ is the momentum fraction of \Q\ relative to the fragmenting parton $k$, $f$ are the collinear Parton Distribution Functions (PDFs), $d\hat{\sigma}$ are the SDCs {in the Zero-Mass VFNS}, and $D$ are the FFs. PDFs and FFs are evaluated at the scales $\mu_f$ and $\mu_F$, respectively, while $\mu_R$ denotes the renormalization scale which enters the strong coupling $\alpha_s$ used for the perturbative expansion of the SDCs.

The NRQCD factorization conjecture~\cite{Bodwin:1994jh}, based on a double expansion in $\alpha_s$ and $v$, states that, at the scale $\mu_{0}\sim M$, FFs can be factorized into calculable SDCs and LDMEs~\cite{Braaten:1993mp,Braaten:1994vv,Braaten:1994kd,Nayak:2005rt}, accounting for the nonperturbative transition of the intermediate heavy-quark-pair state $n={}^{2S+1}L_J^{[C]}$ ($\QQ[n]$) into the physical \Q.\footnote{${}^{2S+1}L_J$ is the usual spectroscopic notation  and $[C]$ is the color state.} The factorized FFs read:
\begin{equation}
\hspace{-0.4em}D_{k/\Q}(z,\mu_{0})\!=\hspace{-0.7em}\sum_{n,m,n_v}\hspace{-0.4em}
\frac{\alpha_s^{m}(\mu_{R_0})}{m^3_Q}
d^{(m,n_v)}_{k/ Q\bar Q[n]}(z,\mu_{R_0},\mu_{0})\,
v^{n_n+n_v}
\langle \widetilde {\cal O}_{n_v}^\Q(n)\rangle,
\label{eq:FF-master}
\end{equation}
where we have exposed the absolute power in $\alpha_s$ as $m$ and the relative order in $v^2$ of each transition as $n_n+n_v$.\footnote{The relative suppression compared to the leading color-singlet (CS) contribution from ${}^3S_1^{[1]}$ is made explicit using  $\langle \widetilde {\cal O}_{n_v}^\Q(n)\rangle = \langle {\cal O}_{n_v}^\Q(n)\rangle /v^{n_n+n_v}$, following the LDME $\langle {\cal O}_{n_v}^\Q(n)\rangle$ definitions of Bodwin~\cite{Bodwin:2012xc}. Note that the notation employed by Ma~\cite{Zhang:2017xoj} differs from that of Bodwin~\cite{Bodwin:1994jh}, used here, by {a factor of} $2N_c$: $\langle \mathcal{O}^\Q_0(^3S_1^{[1]}) \rangle=2 N_c\langle \mathcal{O}^\Q(^3S_1^{[1]}) \rangle_\text{Ma}$.} The first relativistic corrections to ${}^3S_1^{[1]}$ scale like $v^2$ ($n_n=0${, $n_v=2$}). Following the velocity-scaling rules of NRQCD, contributions from Color-Octet (CO) states are suppressed by relative scaling $v^{n_n}$: $n_n=3$ for $^1S_0^{[8]}$, $n_n=4$ for $^3S_1^{[8]}$ and $^3P_J^{[8]}$, etc. Each of these transitions can receive higher relativistic corrections with $n_v=2,4$, and so on. At $\mathcal{O}(v^4)$, the next-to-next-to-leading relativistic corrections to ${}^3S_1^{[1]}$ ($n_n=0$, $n_v=4$) mix with the leading $^3S_1^{[8]}$ and $^3P_J^{[8]}$ contributions ($n_n=4$, $n_v=0$)~\cite{Bodwin:2012xc}, which illustrates that CO contributions are in essence higher{-order} relativistic corrections.

CO LDMEs are not computable from first principles and are usually fitted to data. Yet, fit results often differ by more than an order of magnitude~\cite{Lansberg:2019adr}. Using the Gremm-Kapustin relation~\cite{Gremm:1997dq}, one can connect  the better known CS LDME at relative order $v^0$ to that at relative order $v^2$ up to $\mathcal{O}(v^2)$. 
In our notation, it amounts to using $\langle \widetilde{\cal O}_{2}^{\psip}({}^3S_1^{[1]})\rangle= \langle \widetilde{\cal O}_{0}^{\psip}({}^3S_1^{[1]})\rangle$ and replacing $v^2$ in \ce{eq:FF-master} by its average value  $\langle v^2 \rangle$ {on the order of}  $(M-2m_Q)/m_{Q}$, where $m_Q$ is the heavy-quark mass (see Refs.~\cite{Bodwin:2012xc,Bodwin:2007fz}).\footnote{{Variants of this relation exist, like  $(M^2-4m^2_Q)/(4m^2_Q)$~\cite{Braaten:2002fi} or $(M-2m_Q)/m_{\text{QCD}}$~\cite{Bodwin:2003wh} with $m_{\text{QCD}}$ being the heavy-quark mass in the NRQCD action. In what follows, we fix $m_Q=m_c=1.5$ GeV in that interplays with LDMEs and $\langle v^2 \rangle$ are non trivial}.} Further, by using the vacuum-saturation approximation~\cite{Bodwin:1994jh}, one can relate the CS LDME {for production to that for decay and} to the $\Q$ wave function at the origin, $R(0)$, as $\langle \widetilde{\cal O}_{0}^{\psip}({}^3S_1^{[1]})\rangle= 2 N_c (2J+1) |R(0)|^2/4 \pi$. This makes CS-based computations more predictive than CO-based ones. In what follows, we will thus mainly work up to $\mathcal{O}(v^2)$.

The $\mu_{f,F}$ dependence of PDFs and FFs is governed by the Dokshitzer-Gribov-Lipatov-Altarelli-Parisi (DGLAP) evolution equations~\cite{Gribov:1972ri,Lipatov:1974qm,Dokshitzer:1977sg,Altarelli:1977zs}, which for FFs read:
\begin{equation}
\mu_F^2 \frac{d D_{i/\Q}}{d \mu_F^2}(z,\mu_F) = \sum_j  P_{ji} \left(z, \alpha_s(\mu_F) \right)  \otimes D_{j/\Q}(z,\mu_F),
\label{eq:DGLAP}
\end{equation}
where $P_{ji}$ the are timelike splitting functions. {{\texttt{APFEL++}~\cite{Bertone:2017gds,Bertone:2013vaa} can solve Eq.~(\ref{eq:DGLAP}) at NLO accuracy, thus achieving NLL resummation of collinear logarithms,} with any FF initial conditions in $z$ space (see below). Evolved FFs were tabulated in the \texttt{LHAPDF}  format~\cite{Buckley:2014ana} and benchmarked against \texttt{MELA}~\cite{Bertone:2015cwa}, which performs evolution in Mellin space.}
For the PDFs, we use the CT18NLO set~\cite{Hou:2019qau}. The SDCs $d\hat \sigma$ in Eq.~(\ref{eq:LP}) {are evaluated} at NLO using \texttt{INCNLO1.4}~\cite{Werlen:INCNLO} (which we benchmarked against \texttt{FMNLO}~\cite{Liu:2023fsq}).

We now review what is known of FFs at their initial scale $\mu_0$ in NRQCD for the $^3S_1^{[1]}$ state. 
Contributions to $D_{c (\bar c)/\psip}$ come from $c (\bar c)\to \psip c (\bar c)$ and were first computed at LO $\mathcal{O}(\alpha_s^2)$ for the unpolarized case~\cite{Braaten:1993mp}, then for the polarized\footnote{We refer here to the polar anisotropy in the hadron center-of-mass (or helicity) frame. The entire density matrix is  unknown.}  case~\cite{Bodwin:2014bia}. $\mathcal{O}(\alpha_s^3)$~\cite{Zheng:2019dfk} and $\mathcal{O}(v^4)$ corrections~\cite{Cui:2025wjq,Sang:2009zz} are known in the unpolarized case. 
LO $\mathcal{O}(\alpha_s^3)$ expressions for $D_{g/\psip}$ from $g\to \psip gg$~\cite{Braaten:1993rw} are known analytically~\cite{Zhang:2017xoj}. 
$D_{g/\psip}$ at $\mathcal{O}(v^2)$~\cite{Bodwin:2003wh} and $\mathcal{O}(v^4)$~\cite{Bodwin:2012xc} is known in the unpolarized case and at $\mathcal{O}(v^2)$ in the polarized case~\cite{Zhang:2017xoj}.
{At $\mathcal{O}(\alpha_s^3)$, $g\to \psip c\bar{c}$ also contributes. It has been partially assessed from off-diagonal evolution of $D_{c/\Q}$~\cite{Cacciari:1994dr,Cheung:1993pk} in the 90's. Its full expression can be derived from that of $g\to B^\star_c c\bar{b}$~\cite{Feng:2021qjm} which
 Feng {\it et al.} provided  us for our study.\footnote{All FFs are plotted in the supplementary material.} Up to $\mathcal{O}(\alpha_s^3)$, there is no initial-scale contribution to FFs from
light quarks ($u,d,s$), which are instead generated through evolution.}
In our study, we use the most precise results for each channel.

{\it Phenomenological parameters and benchmarking with earlier results --- } LP $^3S_1^{[1]}$ FF studies made in the 90's~\cite{Roy:1994ie,Braaten:1994xb,Cacciari:1994dr,Braaten:1996pv,Kramer:2001hh}, using both $D_{g/\psip}$ and $D_{c/\psip}$ at LO and evolved separately at LL, found cross sections $\mathcal{O}(30)$ times too small compared to the early Tevatron data~\cite{CDF:1997ykw}: this is known as the CDF $\psip$ ``anomaly'' or ``surplus''. {Using the same setup, we reproduced the results of Refs.~\cite{Kramer:2001hh,Cacciari:1994dr}.} In Ref.~\cite{Braaten:1994xb}, massive SDCs were used for the charm channel. We stress that no systematic studies of the theoretical uncertainty on {$D_{g/\psip}$ and $D_{c/\psip}$ inputs were carried out. 
Since then, our knowledge of PDFs, SDCs, $\alpha_s$, evolution, and $\Q$ parameters has changed. 
In particular, we find that the central value of our LP  $\psip$ cross section increases by a factor of four after having (i) increased perturbative accuracy (SDCs at NLO and resummation at NLL) and (ii) updated input parameters, {i.e.,} the radial wavefunction at the origin $|R(0)|^2$ and  the $\alpha_s$ values.\footnote{We have used $\alpha_s$ from the CT18 (N)LO fits~\cite{Hou:2019qau}, where $\alpha^{1(2)\text{-loop}}_s(M_Z)=0.135(0.118)$, rather than $\alpha^{1\text{-loop}}_s(M_Z)=0.124$, as used in Ref.~\cite{Kramer:2001hh}.} In addition, results are  associated with an uncertainty of at least a factor of five. This comes from the $\mu_{R_0}$ uncertainty on $D_{g/\psip}\propto \alpha_s^3(\mu_{R_0})$ for $\mu_{R_0} \in \{m_c,4m_c\}$, which alone is close to four, since $\alpha_s(m_c)/\alpha_s(4m_c)\simeq1.6$.} We have also used two $J/\psi$ studies at the Tevatron for our benchmarking: one at NLL by Kniehl and Kramer in 1998~\cite{Kniehl:1998qy} and one by Qiao in 2003~\cite{Qiao:2003pu}{. The latter} revisited the impact of $D_{c/J/\psi}$ at LL and found larger cross sections than in Refs.~\cite{Kramer:2001hh,Cacciari:1994dr}. This is {consistent} with our observations and a more recent FO NLO study of $pp\to \psi c \bar c$ by Qiao and Feng~\cite{Feng:2025btt}.

LDMEs (or $|R(0)|^2$) can be obtained from potential models up to $\mathcal{O}(v^4)$. Using the Cornell potential~\cite{Eichten:1979ms}, one finds $|R(0)|^2=0.93$~GeV$^3$~\cite{Eichten:1995ch}. This shifts to $|R(0)|^2=0.70$~GeV$^3$ using its variant~\cite{Eichten:2019hbb} and to $|R(0)|^2=0.53$~GeV$^3$
 using Buchm\"uller-(Grunberg-)Tye potential~\cite{Buchmuller:1980su}. $|R(0)|^2$ can also be extracted from the very well measured leptonic-decay width $\Gamma_{\ell\ell}$. Yet, $\Gamma_{\ell\ell}$ receives large QCD corrections at NLO and even larger at NNLO~\cite{Beneke:1997jm,Czarnecki:1997vz,Beneke:2014qea,Egner:2021lxd,Feng:2022vvk}.
These instabilities preclude using decays to obtain the $\jpsi$ and $\psip$ color-singlet LDMEs.
$\Gamma_{\ell\ell}$ also receives significant relativistic $v^2$ corrections, which reads up to $\mathcal{O}(\alpha_s,v^2)$:
$\Gamma_{\ell\ell}= 4\alpha^2 e_c^2M_{\psip}^{-2} |R(0)|^2(1-16/3 \alpha_s(\mu_R) /\pi - 1/3 \langle v^2\rangle)$ and {tends to} increase $|R(0)|^2$ (see e.g.~\cite{Bodwin:2006yd}).  For instance, for $\alpha_s(\mu_R)=0.3$ and $\langle v^2\rangle =0.5$ and using $\Gamma^\text{exp.}_{\ell\ell}$~\cite{ParticleDataGroup:2024cfk}, one has $|R(0)|^2=1$~GeV$^3$. In fact, even larger values of $|R(0)|^2$ can be obtained with default parameters if the NNLO expression is employed~\cite{ColpaniSerri:2021bla,Huang:2023pmn}.
In view of this uncertainty, we consider three different but realistic values of $|R(0)|^2[\text{GeV}^3]$: $0.50$, $0.75$ and $1.00$.

{\it Impact study of $\mathcal{O}(v^2)$ relativistic corrections --- }Before comparing  data and {$v^2$-improved computations}, we find it important to explain why the large-$z$ region is the most relevant. We use Mellin moments to illustrate this point. Let us first examine the $P_T$ scaling. At {LO}, SDCs scale asymptotically like $d\hat{\sigma_k}\propto P_{T_k}^{-4}$,  where $P_{T_k}= P_T/z$ is {the transverse momentum} of the fragmenting parton $k$. The spectrum {becomes} steeper, {$(z/P_T)^N$} with $N>4$, because of the running of $\alpha_s$ and the convolution with the PDFs. The hadronic cross section is then sensitive to the $N$-th Mellin moment of the FFs: $d\sigma/dP_T\propto \tilde D(N, \mu=P_T) = \int_0^1 dz \,z^{N-1}  D(z, \mu=P_T)$, with $N\simeq 5$ at the LHC  and a little higher at the Tevatron. $z$ values close to one are thus the most important, which is coherent with the idea that a high-$P_T$ hadron is usually produced by a parton which has transferred nearly all of its momentum to it. At LL, decoupled evolution (i.e. neglecting off-diagonal splitting functions) in Mellin space is multiplicative via the factor {$\left[\alpha_s(\mu_F)/\alpha_s(\mu_0)\right]^{-\gamma(N)/\beta_0}$}, where
$\gamma(N)\sim -\ln N$ for large $N$. So a large-$N$ analysis  of the $\mathcal{O}(v^2)$ corrections at $\mu_0$ is in principle sufficient to assess their importance at higher scales. In view of the slower evolution of charm, its importance  grows with $P_T$.

At $\mu_0$, {$\mathcal{O}(v^2)$ corrections to $\tilde{D}_{g\to\psip gg }(N=6.2 ,\mu_0=2m_c)$ are~\cite{Bodwin:2012xc} as large as $\mathcal{O}(15 \langle v^2\rangle)$.} We have Mellin-transformed our results using NLL evolution from $\mu_0$ for charm and gluon FFs and we confirm that this enhancement {holds at higher-scales:\footnote{We gathered evolved Mellin moments for different $N$'s and channels used in this work as supplementary material for the reader's convenience.}
$\tilde D_{g\to\psip gg }(6.2,100$~GeV) gets an $\mathcal{O}(15.6 \langle v^2\rangle)$ correction, while that for $\tilde D_c(6.2,100$~GeV) is $\mathcal{O}(-0.12 \langle v^2\rangle)$. }

At this stage, it is important to comment on realistic values for  $\langle v_{\psip}^2\rangle$. According to the Gremm-Kapustin relation, we expect it to be larger for $\psip$ than for $J/\psi$, yet with large uncertainties, owing to the ambiguity on $m_c$~\cite{Braaten:2002fi}. {As such, we use three values {of  $\langle v_{\psip}^2\rangle$}: 0.25, 0.35, and 0.5.} These values result in an increase of the cross section {from $g\to \psip gg$} at $P_T=100$~GeV by a factor of ranging from 3.8 to 7.5.

%% file: results.tex
For our cross-section evaluations, our default and variation scales are as follows. For $D_g$ and $D_c$, the values of the renormalization scales $\mu^g_{R_0}$ and $\mu^c_{R_0}$ are both defaulted to $2m_c$. The default value of the initial scale for the evolution, $\mu_0$, is also taken to be $2m_c$.\footnote{One could use different values for different channels ($2m_c$ for $D_g$ and $3m_c$ for $D_c$) and use a linearized evolution to pre-evolve some channels to a common $\mu_0$, but{, as long as $\mu_0$ remains in the vicinity of $2m_c$ and $3m_c$,}  results {are almost unaffected}.} The renormalization, initial-state factorization, and final-state factorization scales, $\mu_{R,f,F}$, are all defaulted to $P_T$. Scale variations are performed by varying one scale at a time about its default value by a factor of two ({e.g.}, {$\mu^{\{g,c\}}_{R_0}=\xi^{\{g,c\}}_{R_0} \cdot 2  m_c$  with $\xi^{\{g,c\}}_{R_0}=(0.5,1,2)$}), while setting all other scales to their default. The resulting asymmetric uncertainties are combined in quadrature with PDF uncertainties.

\begin{figure*}[!htbp]\vspace*{-0.25cm}
  \centering 
    \subfigure{%
    \includegraphics[height=6cm]{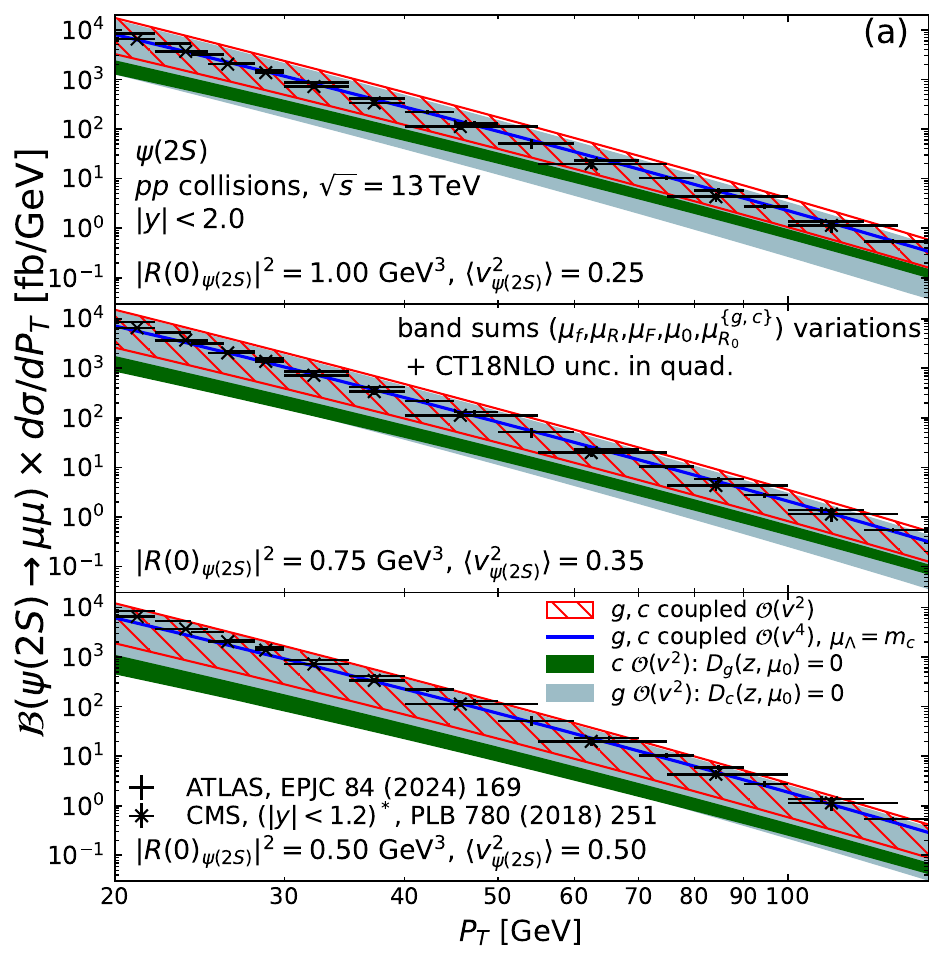}%
     \label{fig:Atlas}
     }\subfigure{%
   \includegraphics[height=6cm]{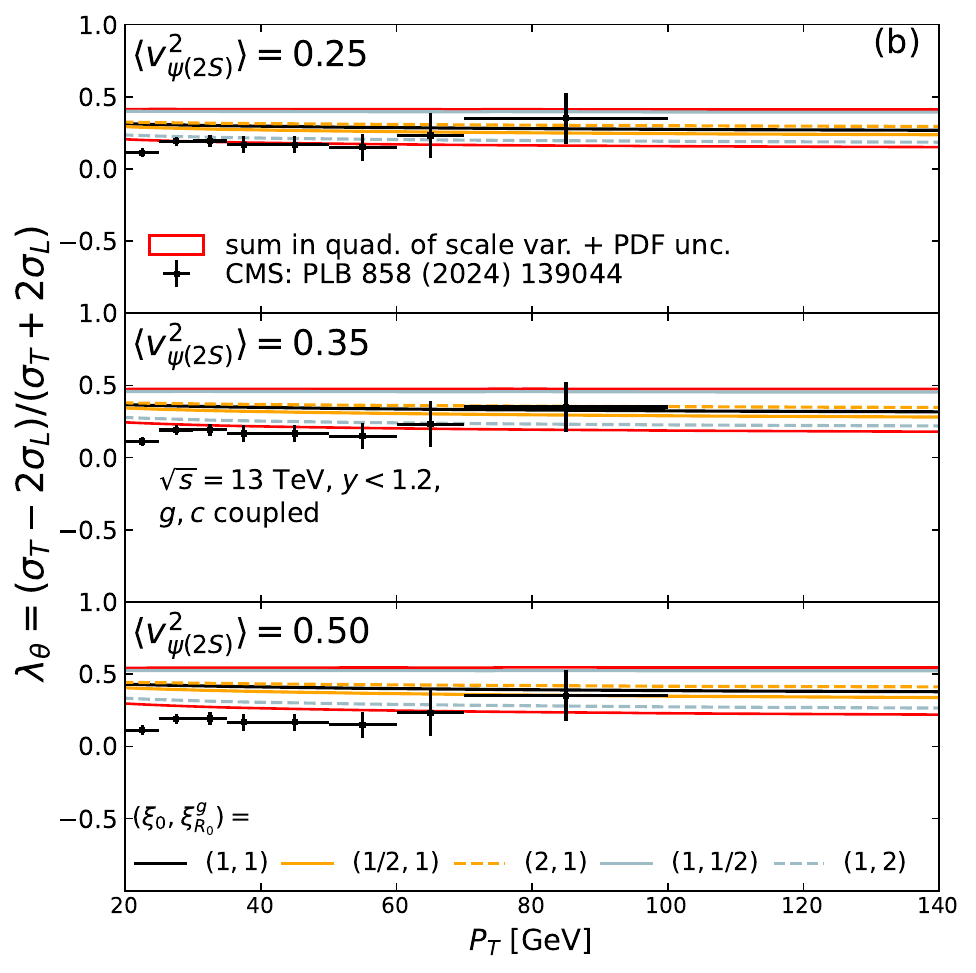}%
   \label{fig:polarisation}
  }\subfigure{%
    \includegraphics[height=6cm]{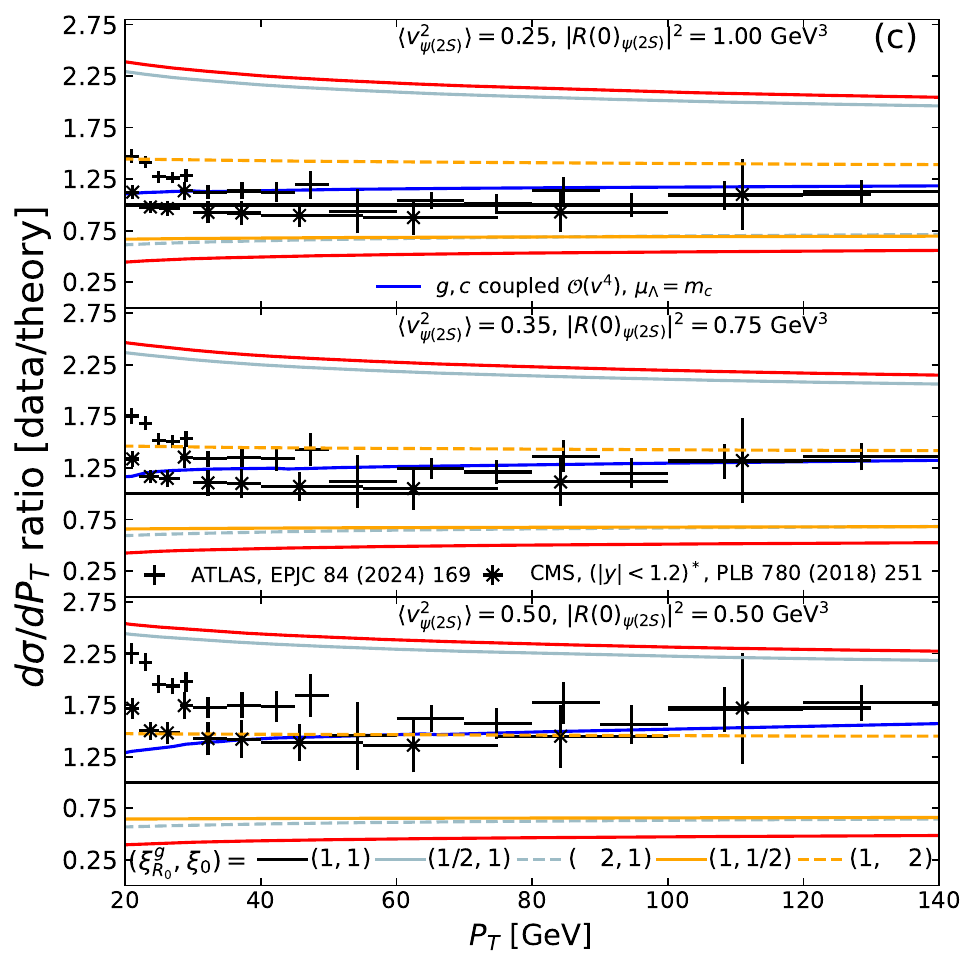}%
    \label{fig:altATLAS}
  }  
\caption{(a) Prompt ATLAS and CMS $\psip$ $P_T$-differential cross sections~\cite{ATLAS:2023qnh,CMS:2017dju} at \(\sqrt{s} = 13~\text{TeV}\) compared to our NLL results for coupled $c$ \& $g$ (hatched red), $c$ (solid green), and $g$ (solid gray-blue) channels. (b) Prompt $\psip$ polar anisotropy $\lambda_\theta$ from CMS~\cite{CMS:2024igk}, compared to our coupled $c$ \& $g$ results. Solid and dashed curves show different scale choices and red curves show the combined scale and PDF uncertainty. (c) data-over-theory ratio with the color code of (b). We also show the $\mathcal{O}(v^4)$ correction (blue line) 
for our central scales and $\mu_\Lambda=m_c$ in (a,c). }
\label{fig:LHC}
\end{figure*}

Fig.~\ref{fig:Atlas} compares the prompt \(\psip\) \(P_T\)-differential data at \(\sqrt{s}=13~\text{TeV}\) in the rapidity region $|y|<2$ measured by ATLAS~\cite{ATLAS:2023qnh} with our NLO+NLL predictions obtained through coupled evolution (hatched red), and separately with decoupled evolution (solid gray-blue for $g$ and solid green for $c$). {We have also plotted the CMS data~\cite{CMS:2017dju} rescaled by the ratio of the rapidity intervals, $\Delta y_\text{ATLAS}/\Delta y_\text{CMS}$.  
Given the correlation between $|R(0)|^2$ and $\langle v^2 \rangle$,
 {we anti-correlated them} to avoid over-inflating uncertainties {choosing: ($|R(0)|^2[\text{GeV}^3],\langle v^2 \rangle)=(1,0.25)$, $(0.75,0.35)$, and $(0.5,0.5)$.} Results are shown in the upper, middle, and lower panels of Fig.~\ref{fig:LHC}, respectively.\footnote{We show additional results for ($|R(0)|^2~[\text{GeV}^3],\langle v^2 \rangle)=(0.5,0.25)$ and $(1.0,0.50)$ in supplementary materials.}
We note that, for each variation, our complete results (hatched red band) agree with data. The main difference lies in  the importance of charm fragmentation (green band) which can amount to up to a third of the yield.} A thorough discussion on the different sources of uncertainty is given below.\footnote{We included a similar figure with predictions for FCC-hh~\cite{FCC:2018vvp} at $\sqrt{s}=100$~TeV for $|y|<2.5$ in the supplementary material.} 

Fig.~\ref{fig:polarisation} compares the latest CMS data~\cite{CMS:2024igk} for the polar anisotropy parameter, $\lambda_\theta\equiv(\sigma_T-2 \sigma_L)/(\sigma_T+2 \sigma_L)$, measured in the hadron helicity frame as a function of $P_T$, with our results for our 3 scenarios. The only difference{ compared to the unpolarized cross section} is that the polarized FFs $D^{T,L}_c$ are only known up to LO in both $\alpha_s$ and $v^2$. Defining the partial cross sections  $\sigma_{\{g,c\}}$ as that from the convolution with $D_{\{g,c\}}$, one knows that $\mathcal{O}(v^2)$ corrections~\cite{Zhang:2017xoj} make $\sigma_g$ more transverse, while one expects $\sigma_c$ to remain unpolarized, as suggested by the recent full  NLO computation of $pp \to \psi c\bar c$~\cite{Feng:2025btt} up to large $P_T$. It follows that a relative increase (decrease) of $\sigma_{g}/\sigma_{c}$ corresponds to $\lambda_\theta$ closer to one (zero).

There is thus a strong correlation between our polarization and cross-section results. To assess it, \cf{fig:altATLAS} shows a data/theory ratio featuring the complete theory uncertainty curves (red) as well as the two largest scale variations (gray-blue for $\mu^g_{R_0}$ and orange for $\mu_{0}$), like in~\cf{fig:polarisation}. Other uncertainties, {e.g.,} those from hard scales ($\mu_{f,F,R}$) and PDFs, are significantly smaller.\footnote{The complete breakdown of the cross-section uncertainties along {with a dedicated plot for charm fragmentation} is given as supplementary material.} {For $\mu^c_{R_0}$, the cross section exhibits a notably reduced sensitivity at NLO~\cite{Zheng:2019dfk}.}

For $\xi_{R_0}^g=1/2$ (solid gray-blue), $\sigma_g > \sigma_c$ and the yield and $\lambda_\theta$ are largest. On the contrary, for $\xi_{R_0}^g=2$ (dashed gray-blue),  $\sigma_c > \sigma_g$ and the yield and $\lambda_\theta$ are smallest. A similar effect is observed for $\mu_{0}$ and follows from the slower evolution of $D_c$. Further conclusions would require {knowing}  $D_g$ at NLO to reduce the $\mu_{R_0}^g$ uncertainty on $\sigma_g$, and of NLO corrections to polarized FFs $D^{L,T}_{c,g}$. At this stage, we limit to note a better agreement for $|R(0)|^2=1\text{GeV}^3$ and $\langle v^2\rangle=0.25$, where charm contributes up to half of the yield. 

{\it Impact study of $\mathcal{O}(v^4)$ relativistic corrections --- }
As just observed, gluon FFs receive large $\mathcal{O}(v^2)$ corrections which align the central values of our results with the LHC experimental data when  charm fragmentation is accounted for.
It is therefore reasonable to question whether the inclusion of $\mathcal{O}(v^4)$ corrections would overshoot the data -- or, equivalently, whether the series in $v^2$ converges.  The CS FFs are known up to $\mathcal{O}(v^4)$~\cite{Bodwin:2012xc,Cui:2025wjq} and enable a NLL study, provided that the distributional $\mathcal{O}(v^4)$ terms are handled in $N$ space (using e.g. MELA).
Yet, it should be noted that a complete $\mathcal{O}(v^4)$ study should include octet channels. This would require refitting existing octet LDMEs, which is beyond the scope of this work. Hence, we limit our exploration {at $\mathcal{O}(v^4)$} to the CS channel by showing a single (blue) curve in  Figs.\ref{fig:LHC} (a) \& (c) for the central scale choice and $\mu_\Lambda=m_c$: the increase (from gluons) is moderate (compare to the black curve) and, in fact, even improves the agreement, particularly 
for $|R(0)|^2=0.5\text{GeV}^3$ and $\langle v^2\rangle=0.5$. We recall that polarized CS FF are not known up to $\mathcal{O}(v^4)$, which prevents us from adding a $\mathcal{O}(v^4)$-improved curve in \cf{fig:polarisation}.

\begin{figure}[!hbp]\vspace*{-0.35cm}
  \centering 
\includegraphics[width=.5\textwidth]{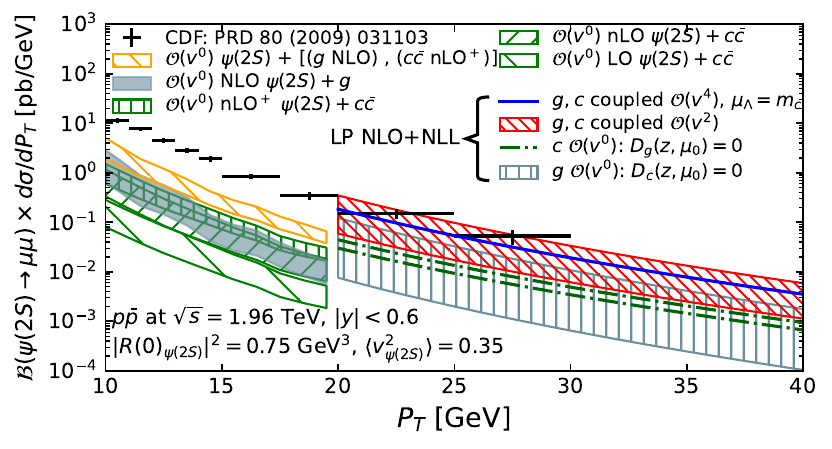}
  \caption{CDF Tevatron data~\cite{CDF:2009kwm} compared, above 20 GeV, to NLO+NLL LP at $\mathcal{O}(v^0, v^2, v^4)$ (same color code as for Fig. 1) and, below 20 GeV, to  {\it approximated} results containing NLPs: nLO=LO$\times K_\text{NLO}$, nLO$^+=$LO$\times K_\text{NLO}\times [\alpha_s(2m_c)/\alpha_s(P_T)]^{3/2}$ -- all at $\mathcal{O}(v^0)$ (see text).
  \label{TeV}}
\end{figure}

{\it Revisiting the CDF $\psi(2S)$ anomaly with NLP corrections---}
Fig.~\ref{fig:altATLAS} shows that the LHC experimental data points below $P_T=30$~GeV begin to deviate from our predictions. This departure can be interpreted as a signal of the onset of next-to-leading power (NLP) corrections, which implies that caution is warranted when comparing our leading-power (LP) results to CDF data at lower $P_T$.
{Fig.~\ref{TeV} shows an estimate of these NLP contributions obtained from existing FO studies valid at mid $P_T$. 
The NLP charm contributions is complicated to assess since $\mathcal{O}(\alpha_s^4)$ contributions from $gg \to \psi(2S) c \bar c $ contain both LP and NLP components~\cite{Artoisenet:2007xi}. At NLO ($\mathcal{O}(\alpha_s^5)$), the yield receives a large ($P_T$-independent) $K$ factor, close to 3.5~\cite{Feng:2025btt}, like for open-charm production.
At higher orders, emissions from the fragmenting charm will be further enhanced by logarithms of $P_T$ which we have roughly estimated by rescaling the yield
by $\left[\alpha_s(P_T)/\alpha_s(2m_c)\right]^{- 2 + \gamma_{qq}(5.7)/ \beta_0 }$.  This recipe happens to match quite well our NLO+NLL LP result around $P_T=20$~GeV and to provide a first estimate of NLP contributions down to $P_T=10$~GeV. This also allows us to note that the charm contribution could be up to a fifth of the observed Tevatron rate. A more precise statement would require matching our LP to the $\mathcal{O}(\alpha_s^5)$ computation, which is beyond the scope of this analysis. We note that the corresponding $\mathcal{O}(\alpha_s^4 v^2)$ corrections are unknown but, unlike the gluon case (see below), we have no reason to believe that they could be large.}
{For the gluons, the situation is simpler as NLO $\mathcal{O}(\alpha_s^4 v^0)$ corrections are at most NLP~\cite{Campbell:2007ws}. In principle, they can be simply summed to our LP results without any double counting. They are shown in gray-blue. As expected, one observes an increase from the $P_T^{-6}$ contributions at $P_T=10$~GeV where $\mathcal{O}(\alpha_s^4 v^0)$ NLP gluon contributions could be up to a third of the measured rate.
Yet, while $\mathcal{O}(\alpha_s^3 v^2)$ corrections were found in 2009~\cite{Fan:2009zq} to be  ${\cal O}(1/6 \langle v^2\rangle$), $\mathcal{O}(\alpha_s^4 v^2)$ corrections to NLPs are unknown and we cannot extrapolate our $\mathcal{O}(v^2)$ LP results to lower $P_T$ values. At this stage, our observations in this region are only indicative and call for a specific effort in augmenting existing FO computations with $\mathcal{O}(v^2)$ corrections.}

%% file: Conclusion.tex
We have studied the impact of {${\cal O}(v^2)$} relativistic corrections to $\psip$ production. We have used all current knowledge on quarkonium FFs at the initial scale and performed evolution at NLL with SDCs computed at NLO. We have performed a thorough investigation of theoretical uncertainties, from PDFs, FFs, and  SDCs. Those from FFs are the largest, in particular from the renormalization scale of the initial-scale gluon FF, underscoring the need for a NLO determination of the latter. ${\cal O}(v^2)$ corrections boost the $g\to \psip gg$ contribution by a factor 3.8 and 7.5 depending on the value of $\langle v^2 \rangle$. {The charm contributions are non-negligible and stable.} The resulting central values of our NLO+NLL LP results agree with the latest ATLAS and CMS cross-section data without the need to include higher-order relativistic corrections from, e.g., CO transitions. Our results also approach consistency with recent polarization measurements by CMS {partially because of the presence of charm fragmentation}. We note that the unknown NLO corrections to the gluon channel are likely to produce depolarization, which would further improve the agreement. 
Our results also underscore the need to improve our understanding of NLP corrections in the region $P_T < 30$~GeV, which is needed to address the Tevatron data. At larger $P_T$, besides the computation of the NLO correction to the (polarized) gluon FF, one should investigate  how these large ${\cal O}(v^2)$ CS corrections will impact existing ${\cal O}(v^{3,4})$ CO LDME determinations~\cite{Butenschoen:2022qka,Bodwin:2015iua,Shao:2014yta,Gong:2012ug}.